\documentclass[journal]{IEEEtran}
\usepackage[utf8]{inputenc}
\usepackage{amsmath}
\usepackage{amsfonts}
\usepackage{amsmath}
\usepackage{amssymb}
\usepackage{array}
\usepackage{bm}
\usepackage{cases}
\usepackage{color}
\usepackage{enumitem}
\usepackage[dvipsnames]{xcolor} 
 
\usepackage[overload]{empheq}
\usepackage{graphicx}   
\usepackage{makecell}
\usepackage{mathrsfs}
\usepackage{multirow} 
\usepackage{threeparttable} 

\usepackage{physics}    
\usepackage{soul}

\usepackage{subcaption}
\usepackage{xspace}

\usepackage{url}
\usepackage[all]{xy}
\usepackage[font=footnotesize,labelsep=period]{caption}

\newcommand{\PreserveBackslash}[1]{\let\temp=\\#1\let\\=\temp}
\newcolumntype{C}[1]{>{\PreserveBackslash\centering}p{#1}}
\newcolumntype{R}[1]{>{\PreserveBackslash\raggedleft}p{#1}}
\newcolumntype{L}[1]{>{\PreserveBackslash\raggedright}p{#1}}

\definecolor{revAcolor}{HTML}{FFF68F} 
\definecolor{revBcolor}{HTML}{98FB98}
\definecolor{revCcolor}{HTML}{B0E2FF}
\definecolor{revDcolor}{HTML}{FFB6C1}

\makeatletter
\def\subsubsection{\@startsection{subsubsection}
                                 {3}
                                 {\z@}
                                 {0ex plus 0.1ex minus 0.1ex}
                                 {0ex}
                                 {\normalfont\normalsize\itshape}}
\makeatother

\usepackage{tikz}   
\usepackage{standalone}
\usepackage{pgfplots}
\usetikzlibrary{shapes, arrows,chains,calc,shapes}
\pgfplotsset{compat = newest}
\pgfdeclarelayer{bg}    
\pgfsetlayers{bg,main}  

\def\insys{\textit{InSys}\xspace}
\def\exsys{\textit{ExSys}\xspace}
\def\neudyn{{NeuDyE}\xspace}
\def\odenet{{ODE-Net}\xspace}
\def\pineudyn{{PI-NeuDyE}\xspace}
\def\pgneudyn{{PG-NeuDyE}\xspace}

\usepackage{cite}
\graphicspath{ {fig/} }
\title{Physics-Aware Neural Dynamic Equivalence of Power Systems}

\author{Qing~Shen, Yifan~Zhou, Qiang~Zhang, Slava~Maslennikov, Xiaochuan~Luo, and Peng~Zhang
\thanks{
This work was supported in part by the U.S. Department of Energy’s Office of Energy Efficiency and Renewable Energy (EERE)  under the Solar Energy Technologies Office Award Number 38456 
and in part by the National Science Foundation under Grant No. OIA-2134840. 

Q. Shen, Y. Zhou and P. Zhang are with the Department of Electrical and Computer Engineering, Stony Brook University, NY, USA (e-mail: yifan.zhou.1@stonybrook.edu). Q. Zhang, S. Maslennikov and X. Luo are with ISO New England Inc., Holyoke, MA 01040 USA.} 
}

\begin{document}

\maketitle

\begin{abstract}
This letter devises Neural Dynamic Equivalence (\neudyn), which explores physics-aware machine learning and neural-ordinary-differential-equations (\odenet) to discover a dynamic equivalence of external power grids while preserving its dynamic behaviors after disturbances. 
The contributions are threefold:
(1) an \odenet-enabled \neudyn formulation to enable a continuous-time, data-driven dynamic equivalence of power systems;
(2) a physics-informed \neudyn learning method (\pineudyn) to actively control the closed-loop accuracy of \neudyn without an additional verification module;
(3) a physics-guided \neudyn (\pgneudyn) to enhance the method's applicability  even in the absence of analytical physics models. 
Extensive case studies in the NPCC system validate the efficacy of \neudyn, and, in particular, its capability under various contingencies. 
\end{abstract}
\vspace{-3pt}

\vspace{-8pt}
\section{Introduction}
\label{sec:intro}
\IEEEPARstart
{D}{iscovering} reliable dynamic equivalent models of unidentified subsystems or external systems is of critical significance for the resilient operations of large-scale interconnected power systems~\cite{ourari2006dynamic}. However, it is a long-standing obstacle due to the strongly nonlinear dynamics of power systems~\cite{acle2019parameter}, complicated coherency characteristics~\cite{tyuryukanov2020slow}, unavailable component models, etc. 
Recently, the wide adoption of PMUs and the high-rate measurement streams generated from them stimulate the development of data-driven dynamic equivalence. 
While different attempts have been reported, two major challenges remain: 
I) How to retain the continuous-time dynamic behaviors of dynamic equivalence using discrete-time measurements? II) How to theoretically guarantee the closed-loop performance of dynamic equivalence to support its co-simulation in the entire interconnected power systems?


To bridge the gap, this letter devises Neural Dynamic Equivalence (\neudyn). The key innovation is the integration of an \odenet-enabled dynamic equivalence model and a physics-aware \neudyn learning to discover a \textit{continuous-time} dynamic equivalence of power grids with explicitly guaranteed \textit{closed-loop dynamic behaviors} under disturbances. 

\vspace{-8pt}
\section{\neudyn Formulation via \odenet}\label{sec:neudye formulation}

Denote the subsystem to be equivalenced as the external system (\exsys) and the rest as the internal system (\insys). 


Considering the {continuous-time} dynamic natures of power grids, we formulate the dynamic equivalence of \exsys by a set of neural ordinary differential equations (\odenet\cite{NEURIPS2018}):

\noindent
\begin{equation}
\vspace{-5mm}
    \Dot{x}_{ex} = \mathcal{N}_{\theta}({x}_{ex} , s_{in}) \label{equ:exsys model}
\end{equation}
where ${x}_{ex}$ denotes the selected \exsys's dynamic states; $s_{in}$ denotes \insys's impact on \exsys; $\mathcal{N}_{\theta}$ is an ODE-Net parameterized by $\theta$.
Consequently, the entire power grid integrating \exsys and \insys appears as a physics-neural hybrid system:
\begin{subequations}\label{equ:model:whole}
\begin{align}[left = \empheqlbrace\,]
    & \Dot{x}_{in}  = \mathcal{P} ({x}_{in} , {y}_{in} , y_{b} ) \label{equ:model:whole:1}\\
    & \Dot{x}_{ex} = \mathcal{N}_{\theta}({x}_{ex} , s_{in}) \label{equ:model:whole:2}\\
    & \mathcal{G}({x}_{in} , {x}_{ex} , {y}_{in} , y_{b} )= 0 \label{equ:model:whole:3}
\end{align}
\end{subequations}
where ${x}_{in}$ and ${y}_{in}$ denote the dynamic and algebraic states of \insys; $y_{b}$ denotes the boundary states between \insys and \exsys; $\mathcal{P}$ and $G$ denote the dynamic and algebraic models of \insys, which are readily obtained from the \insys physics. 

Obviously, the dynamic simulation of \eqref{equ:model:whole} requires assembling the \insys physics model and  the \exsys neural model, which is referred to as the \textit{closed-loop simulation} of \insys and \exsys in the following discussion.


\vspace{-8pt}
\section{Physics-Aware \neudyn}\label{sec:pi}

\vspace{-2pt}
\subsection{Data-Driven Training for \neudyn and Deficiency Analysis}\label{sec:odenet}

Ideally, \eqref{equ:exsys model} can be independently trained as follows~\cite{zhouyf}: 
\begin{equation}\label{equ:ODENet:loss min:1}
    \min_{{\theta}} \sum\nolimits_i  \| {x}_{ex,i} -  \hat{{x}}_{ex,i} \|_2  ~~,~~ s.t.~~ \Dot{x}_{ex} = \mathcal{N}_{\theta}({x}_{ex} , \hat{s}_{in} )
\end{equation}
where $i$ denotes time point $t_i$; \textsuperscript{$\wedge$} denotes the measurements.

Equation \eqref{equ:ODENet:loss min:1} minimizes the error between the measured \exsys states and the solution of \odenet (i.e., \eqref{equ:model:whole:2}).
An obvious issue  is that it only considers the impact from \insys on \exsys (i.e., \eqref{equ:model:whole:2}) while the impact from  \exsys on \insys (i.e., \eqref{equ:model:whole:1}) is ignored, leading to an \textit{open-loop training} without involving the \textit{closed-loop simulation}. Such an open-loop manner fails to explicitly monitor and control the closed-loop accuracy of \eqref{equ:model:whole} in its training process. 
Actually, even a tiny numerical error in the \exsys model can perturb the \insys dynamics (see \eqref{equ:model:whole:1}). When such errors accumulate during the interactions between \exsys and \insys, the dynamic equivalence may potentially lose efficacy. 


\vspace{-8pt}
\subsection{Physics-Informed Learning for \neudyn}\label{sec:pi:training}
To bridge the gap, we devise a physics-informed \neudyn (\pineudyn), which actively controls the closed-loop accuracy of \neudyn through a physics-aware training process.

The principal idea is to leverage the well-established physics laws of \insys to assist the training of \neudyn for \exsys.
Accordingly, the following training model is developed:

\begin{subequations}\label{equ:pi:model}
\begin{align}
\min_{{\theta}} \sum_{i=1}^n  L_i 
& {=} \sum_{i=1}^n   \| {x}_{ex,i} {-}  \hat{{x}}_{ex,i} \|_2{+}
\|{x}_{in,i} {-}  \hat{{x}}_{in,i} \|_2  \label{equ:pi:model:1}\\
s.t.~~ \Dot{x}_{in} & = \mathcal{P} ({x}_{in} , {y}_{in} , y_{b} ) ~~,~~\Dot{x}_{ex} = \mathcal{N}_{\theta}({x}_{ex} , s_{in} ) \label{equ:pi:model:3}
\end{align}
\end{subequations}

\begin{figure}[!ht]
\vspace{-12pt}
    \centering
     \begin{subfigure}{0.49\columnwidth}
        \includegraphics[width=\columnwidth]{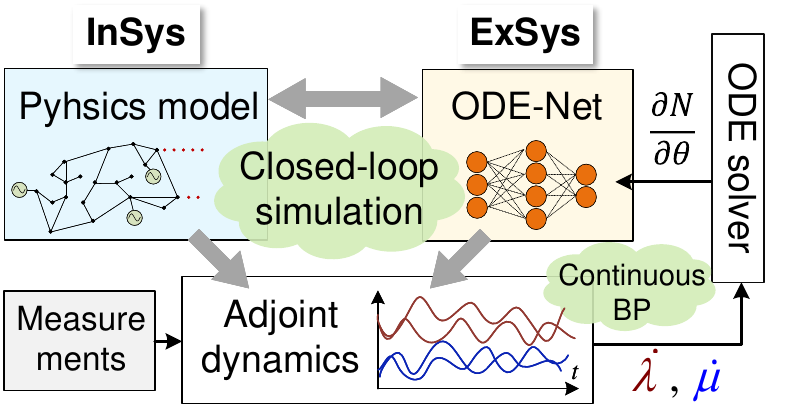}
        \caption{Physics-informed (PI) \neudyn}
         \label{fig1:1}
    \end{subfigure}
    \begin{subfigure}{0.49\columnwidth}
        \includegraphics[width=\columnwidth]{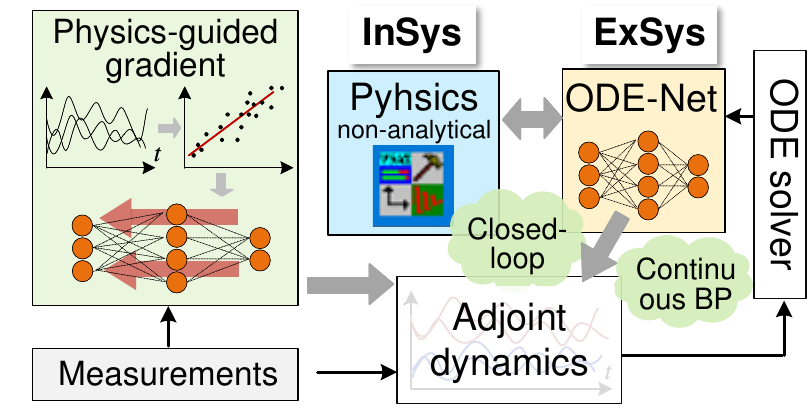}
        \caption{Physics-guided (PG) \neudyn}
        \label{fig1:2}
    \end{subfigure}
    \vspace{-3pt}
    \caption{Schematic diagram of physics-aware \neudyn. } \label{fig1}  
\vspace{-12pt}
\end{figure}
 
As illustrated in Fig.~\ref{fig1:1}, different from \eqref{equ:ODENet:loss min:1}, the model in \eqref{equ:pi:model} explicitly embeds the accuracy of both \exsys and \insys states in a closed-loop manner, which ensures \neudyn generates dependable dynamic responses in conformance with the system's real dynamics once the training converges.

Meanwhile, the training model in {\eqref{equ:pi:model}} differentiates from conventional discrete-time learning because it 
directly minimizes the difference between real and trained dynamic states without any discretization.
This continuous-time learning manner is theoretically more invulnerable to residue training errors and non-ideal measurements, which will be further illustrated in Subsection~{\ref{sec:sim:continuous}}.

Training the model in {\eqref{equ:pi:model}} is not trivial. The major difficulty lies in the fact that ODE-Net's output is the derivative of $x_{ex}$ (see \mbox{\eqref{equ:pi:model:3}}), but the objective targets minimizing the error between $x_{ex}$ and $\hat{x}_{ex}$ (see \mbox{\eqref{equ:pi:model:1}}). 
Therefore, the conventional gradient descent can not be directly applied.
To address the challenge, a physics-informed continuous backpropagation (BP) technique is developed to optimize \eqref{equ:pi:model}. 
We leverage the adjoint method~\cite{NEURIPS2018} to handle the differential constraints in \eqref{equ:pi:model:3}:
\begin{equation}\label{equ:ODENet_PDI:loss}
\begin{aligned}
    \mathcal{L} {=} & \sum_{i=1}^n L_i {-} \int_{t_0}^{t_n} 
    \left[{\lambda}^T (\Dot{x}_{ex} {-} \mathcal{{N}}_{\theta} )  + {\mu}^T ( {x}_{in} {-} \mathcal{\Tilde{P}}) \right] \dd t
\end{aligned}
\end{equation}
where ${\lambda}$ and ${\mu}$ respectively denote the adjoint states for \exsys and \insys; $\mathcal{\Tilde{P}}$ is equivalently reformulated from $\mathcal{P}$ by incorporating \eqref{equ:model:whole:3}, leading to a function of $x_{ex}$ and $x_{in}$.

Accordingly, the gradient of  $\mathcal{L}$ w.r.t $\theta$ is calculate as:
\begin{small}
\begin{equation}\label{equ:pi:L function}
\begin{aligned}
    &\pdv{\mathcal{L}}{{\theta}} = 
    \sum_{i=1}^n \left( \pdv{ L_i }{ {x}_{ex,i}} \pdv{ {x}_{ex,i} }{ {\theta} }+\pdv{ L_i }{ {x}_{in,i} } \pdv{ {x}_{in,i} }{ {\theta} }\right)-    \\
  &\sum_{i=1}^n \int_{t_{i-1}}^{t_i} {\lambda}^T \left(\pdv{\Dot{{x}}_{ex}}{{\theta}} - \pdv{ \mathcal{N} }{ {x}_{ex} } \pdv{ {x}_{ex} }{ {\theta} } -\pdv{ \mathcal{N} }{ {x}_{in} } \pdv{ {x}_{in} }{ {\theta} }- \pdv{ \mathcal{N} }{ {\theta} }  \right)\dd t  \\
    &- \sum_{i=1}^n \int_{t_{i-1}}^{t_i} {\mu}^T \left(\pdv{\Dot{{x}}_{in}}{{\theta}} - \pdv{ \mathcal{\Tilde{P}} }{  {x}_{in} } \pdv{  {x}_{in} }{ {\theta} }-\pdv{ \mathcal{\Tilde{P}} }{ {x}_{ex} } \pdv{ {x}_{ex} }{ {\theta} }  \right) \dd t
\end{aligned}
\end{equation}
\end{small}

With proper adjoint boundary conditions~\cite{zhouyf}, the physics-informed gradient can be yielded from \eqref{equ:pi:L function}, which includes the ``adjoint dynamics'' of ${\lambda}$ and ${\mu}$ (see \eqref{equ:pi:L dynamics:1} and \eqref{equ:pi:L dynamics:2}) and the ``gradient dynamic'' for $\pdv*{\mathcal{L}}{{\theta}}$ (see \eqref{equ:pi:L dynamics:3}):
\begin{subequations}\label{equ:pi:L dynamics}
\begin{align}
 &\dv{{\lambda}^T}{t} = - {\lambda}^T \pdv{ \mathcal{N} }{ {x}_{ex} }- {\mu}^T \pdv{ \mathcal{\Tilde{P}} }{ {x}_{ex} }\label{equ:pi:L dynamics:1}\\
&\dv{{\mu}^T}{t} =- {\lambda}^T \pdv{ \mathcal{N} }{ {x}_{in} } - {\mu}^T \pdv{ \mathcal{\Tilde{P}} }{ {x}_{in} } \label{equ:pi:L dynamics:2}\\
&\dv{}{t} \left( \pdv{\mathcal{L}}{{\theta}}  \right)
= {\lambda}^T \pdv{ \mathcal{N} }{ {\theta} } \label{equ:pi:L dynamics:3}
    \end{align}
\end{subequations}
Finally, the gradient descent for $\mathcal{N}_{\theta}$ can be performed using  $\pdv*{\mathcal{L}}{{\theta}} |_{t=0}$ integrated from \eqref{equ:pi:L dynamics} by arbitrary ODE solvers.

Salient features of \pineudyn over previous purely data-driven learning~\cite{NEURIPS2018,zhouyf} are twofold: 
I) Via the physics-informed model in \eqref{equ:pi:model}, \pineudyn theoretically ensures the \textit{continuous-time} dynamic behaviors of both \insys and \exsys align with discrete-time measurements.
II) Via the physics-informed gradient descent in \eqref{equ:pi:L dynamics}, \pineudyn explicitly controls the closed-loop accuracy of the dynamic equivalence.

In this paper, $\mathcal{N}_{\theta}$ is  constructed by fully-connected neural networks. However, \mbox{\neudyn} can flexibly incorporate arbitrary more advanced neural network structures~\mbox{\cite{qin}\cite{Li2022OnLT}}.

\vspace{-15pt}
\subsection{Physics-Guided \neudyn}\label{sec:pg}

One complication of \pineudyn is that it requires the gradient of \insys's physics models in the training process, i.e., $ \pdv{ \mathcal{\Tilde{P}} }{ {x}_{ex} }$ and $ \pdv{ \mathcal{\Tilde{P}} }{ {x}_{in} }$ in \eqref{equ:pi:L dynamics}. 
In some applications, the analytical expression of those gradients may not be accessible, for example, when \insys is modeled in commercial software such as TSAT or PSS/E. 
Therefore, we further develop a physics-guided \neudyn (\pgneudyn). As illustrated in Fig.~\ref{fig1:2}, it leverages available \insys measurements and grid sparsity to estimate the gradient in the absence of analytical \insys models.

According to the Taylor expansion, the dynamic equations of \insys (i.e., \eqref{equ:model:whole:1}) can be reformulated as:
\begin{equation}\label{equ:pg:taylor}
 {\mathcal{\Tilde{P}}}^{(m)}_i = {\mathcal{\Tilde{P}}}^{(0)} {+}
 [\pdv{ \mathcal{\Tilde{P}} }{ {x}_{ex} } , \pdv{ \mathcal{\Tilde{P}} }{ {x}_{in} }]
\begin{bmatrix}
\hat{x}_{ex,i}^{(m)} - \hat{x}_{ex}^{(0)}\\ \hat{x}_{in,i}^{(m)} - \hat{x}_{in}^{(0)}
\end{bmatrix}
    {+} O((\Delta x)^2)
\end{equation}
where $\hat{x}^{(m)} = [\hat{x}_{ex}^{(m)};\hat{x}_{in}^{(m)}]$ denotes the $m^{th}$ set of time-series measurements of \exsys and \insys at time point $t_i$; $\hat{x}^{(0)} = [\hat{x}_{ex}^{(0)};\hat{x}_{in}^{(0)}]$ denotes a typical operating point (e.g., an equilibrium point);
${\mathcal{\Tilde{P}}}^{(m)}_i = {\mathcal{\Tilde{P}}}(\hat{x}_{ex,i}^{(m)},\hat{x}_{in,i}^{(m)})$; ${\mathcal{\Tilde{P}}}^{(0)} = {\mathcal{\Tilde{P}}}(\hat{x}_{ex}^{(0)},\hat{x}_{in}^{(0)})$.


Denote $A$ as an estimation of $ [\pdv{ \mathcal{\Tilde{P}} }{ {x}_{ex} } , \pdv{ \mathcal{\Tilde{P}} }{ {x}_{in} }]$. 
While the exact elements of $A$ are unknown, the sparsity structure of $A$ is usually accessible due to the connection information in power grids. 
Hence, the $k$-th row of $A$, i.e., $A_k$, can be estimated by the least square regression using available data samples:
\begin{equation}\label{equ:pg:Ak}
    \mathcal{S}(A_k^T) \approx \left(\mathcal{S}(X_k) \mathcal{S}(X_k)^T \right)^{-1}\mathcal{S}(X_k) p_k
\end{equation}
Here, $\mathcal{S}(\cdot)$ denotes a sparsity transformation function, where $\mathcal{S}(A_k^T)$ extracts all non-zero elements of $A_k^T$ 
and $\mathcal{S}(X_k)$ extracts the corresponding elements of $X_k$. $X_k$ and $p_k$ are derived from the trapezoidal rule of \eqref{equ:pg:taylor}, which respectively gather the $k$-th row of  $\hat{x}_{i}^{(m)}+\hat{x}_{i-1}^{(m)} {-} 2\hat{x}^{(0)}$ and of the remained terms of \eqref{equ:pg:taylor} for arbitrary $m$ and $i$. 



Accordingly, an estimation of $A$ can be recovered from $ \mathcal{S}(A_k^T)~(\forall k)$, and, in this way, supports the implementation of the physics-informed gradient calculation in \eqref{equ:pi:L dynamics} if the analytical gradients are unavailable.



\vspace{-8pt}
\section{Case Study}\label{sec:sim}
\begin{figure}[!ht]
\vspace{-10pt}
    \centering
    \includegraphics[width=0.85\columnwidth]{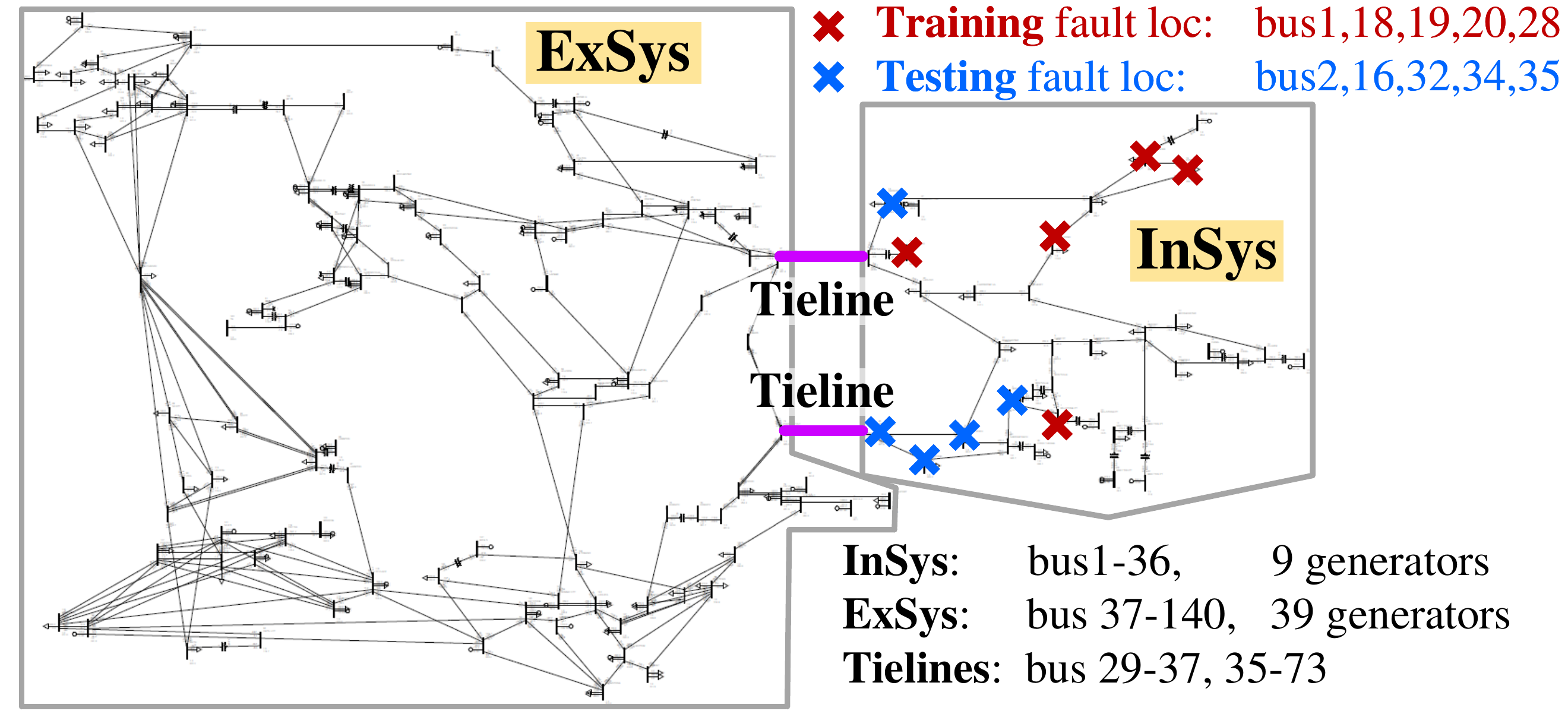}
    \caption{Topology of the NPCC system and fault scenario settings}
    \label{NPCC}
    \vspace{-12pt}
\end{figure}

This section performs case studies on the Northeast Power Coordinating Council (NPCC) system (see Fig.~\ref{NPCC}). The New England system (i.e., buses 1-36) is considered as the \insys, and the rest is the \exsys to be learned by \neudyn. All codes are developed and implemented in Matlab R2022b.
\vspace{-14pt}
\subsection{Experiment Settings}
As shown in Fig.~\ref{NPCC}, 20 training scenarios are generated by  short-circuits 
occurring at bus 18, 19, 20, 21, or 28 with different fault clearing times. 
108 testing scenarios are generated with new fault locations and random fault clearing times at bus 2, 5, 9, 16, 25, 28, 32, 34 and 35.

The electromechanical simulations of the NPCC system are performed via the Power System Toolbox (PST). 27 generators are formulated by the electromechanical model and 21 generators are formulated by the voltage-behind-transient-reactance model, which align with the Transient Security Assessment Tool (TSAT) model provided by ISO NE. %
The trapezoidal rule is adopted for the numerical integration of {\neudyn}.

To learn the \mbox{\neudyn} model of {\exsys} shown in {\eqref{equ:exsys model}}, this paper selects the states of generators, exciters, governors, and line currents of \mbox{\insys} as $s_{in}$ (i.e., features of \mbox{\insys}), and the tieline currents as $x_{ex}$ (i.e., the states of \mbox{\exsys}). However, such training features can be flexibly adjusted according to available measurements.


\vspace{-10pt}
\subsection{Efficacy of Physics-Informed \neudyn}\label{sec:sim:PI}
\subsubsection{Accuracy}
\begin{figure}[!ht]
   \vspace{-18pt}
     \centering
     \includegraphics[width=\columnwidth]{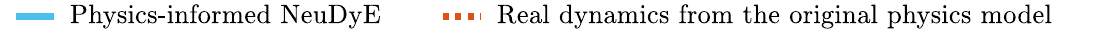}
     \begin{subfigure}{\columnwidth}
        \includegraphics[width=0.48\columnwidth]{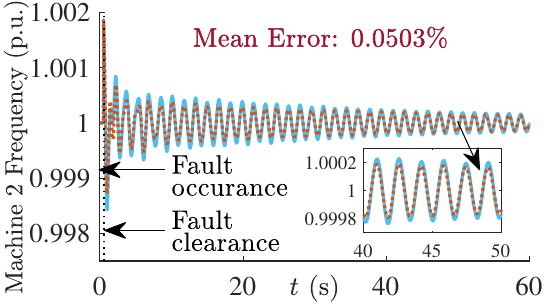}
        \includegraphics[width=0.48\columnwidth]{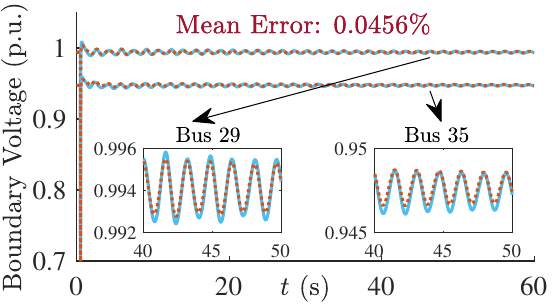}
         \vspace{-3pt}
        \caption{Fault cleared at 0.5400s }
         \label{fig:PI trajectory:shortFCT}
    \end{subfigure}
    \begin{subfigure}{\columnwidth}
        \includegraphics[width=0.48\columnwidth]{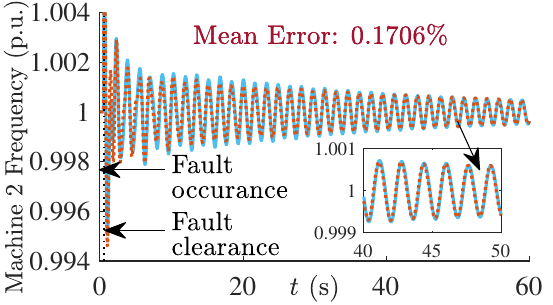}
        \includegraphics[width=0.48\columnwidth]{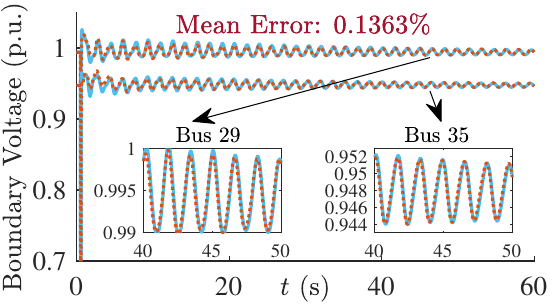}
         \vspace{-3pt}
        \caption{Fault cleared at 0.6108s }
        \label{fig:PI trajectory:longFCT}
    \end{subfigure}
    \vspace{-3pt}
     \caption{Accuracy of \pineudyn under a trained fault location (i.e., bus 21).
     }
    \label{fig:PI trajectory}
    \vspace{-10pt}
\end{figure}

We first validate the accuracy of \pineudyn. 
Fig.~\ref{fig:PI trajectory} presents the performance of \pineudyn under short-circuit faults at bus 21 (a trained fault location) but with random fault clearing times. Trajectories of boundary bus voltages (i.e., bus 29 and bus 35) and machine frequencies of the New England grid  show a perfect match between \neudyn results (see blue lines) and real NPCC dynamics (see red lines), which illustrates the accuracy of the developed method. 

Another observation from {Fig.~\ref{fig:PI trajectory}} is the low-damping oscillation, which is induced by the inter-area modes of the NPCC system. The efficacy of {\neudyn} in capturing both the fast oscillations and the slow damping tendency demonstrates its applicability in stiff systems with multi-time-scale dynamics.

\subsubsection{Generalization capability}
\begin{figure}[!ht]
\vspace{-8pt}
 \centering
 \includegraphics[width=\columnwidth]{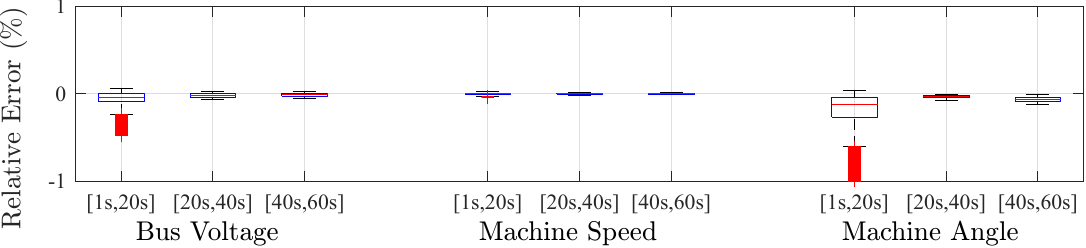}
  \vspace{-15pt}
  \caption{{Accuracy of \pineudyn under new fault locations.}}
     \label{fig:PI generalization}
 \end{figure}

Fig.~\ref{fig:PI generalization} quantitatively studies the generalization capability of \pineudyn with new fault locations and fault clearing times. The time-series relative error of typical system states shows that even under unforeseen faults occurred at new locations, \pineudyn maintains reasonable error rates along the time horizon, which indicates the efficacy of \pineudyn to preserve the dynamic behaviors of \exsys after contingencies and its satisfactory generalization capability beyond the training set.

 \vspace{-10pt}
\subsubsection{Performance in parametric cases}
\begin{figure}[!ht]
\vspace{-10pt}
\centering
\includegraphics[width=\columnwidth]{legend_PDI.pdf}
\includegraphics[width=0.5\columnwidth]
      {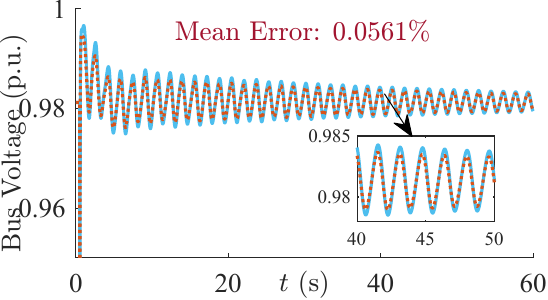}
       \vspace{-3pt}
\caption{Efficacy of PI-NeuDyE in parametric systems.} \label{fig:load}
\vspace{-10pt}
\end{figure}

We further demonstrates the efficacy of \mbox{\pineudyn} under system parameter changes. Without loss of generality, load change is selected as an impact factor to retrain the ODE-Net model, where  the load level of \mbox{\insys} randomly changes from 70\% to 130\% of the original load level.
\mbox{Fig.~\ref{fig:load}} presents the test results of a specific case where the power load increases by 28\% in fault conditions. The relatively low mean error illustrates the accuracy of the method in parametric dynamic systems. 


\vspace{-15pt}
\subsection{Validity of Physics-Guided \neudyn}
\begin{figure}[!ht]
\vspace{-8pt}
\centering
\includegraphics[width=\columnwidth]{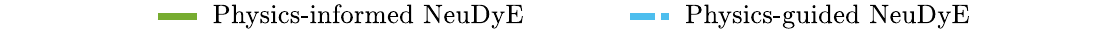}
     \begin{subfigure}{0.47\columnwidth}
        \includegraphics[width=\columnwidth]{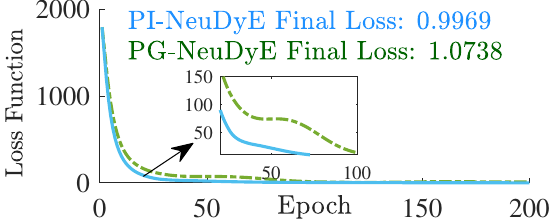}
         
        \caption{Loss function evolution comparison}
         \label{fig:PG:loss}
    \end{subfigure}
    \begin{subfigure}{0.47\columnwidth}
        \includegraphics[width=\columnwidth]{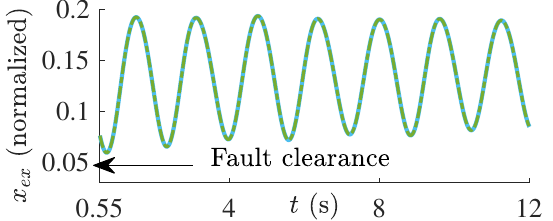}
        
        \caption{Dynamic trajectory comparison}
        \label{fig:PG:traj}
    \end{subfigure}
    \vspace{-3pt}
     \caption{Performance of \pgneudyn and its comparison with \pineudyn.}
    \label{fig:PG}
    \vspace{-10pt}
\end{figure}

Fig.~\ref{fig:PG} presents the performance of \pgneudyn. As shown in Fig.~\ref{fig:PG:loss}, although \pgneudyn converges slightly slower than \pineudyn in the training process (because it does not use any information from the analytical physics models of \insys), it finally reaches a comparable loss level compared with \pineudyn. Meanwhile, Fig.~\ref{fig:PG:traj} shows that the dynamic equivalence obtained from \pgneudyn produces nearly identical dynamic behaviors with that from \pineudyn, which indicates the validity of \neudyn. 

\vspace{-15pt}
\subsection{Comparison with Existing Methods}
Finally, this subsection compares the devised method with existing methods to reveal its necessity and superiority.

\subsubsection{Necessity of the continuous-time consideration}\label{sec:sim:continuous}
\begin{figure}[!ht]
\vspace{-10pt}
    \centering
    \includegraphics[width=\columnwidth]{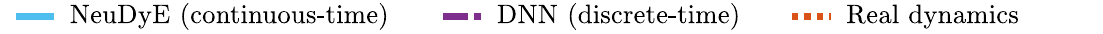}
     \begin{subfigure}{0.47\columnwidth}
        \includegraphics[width=\columnwidth]{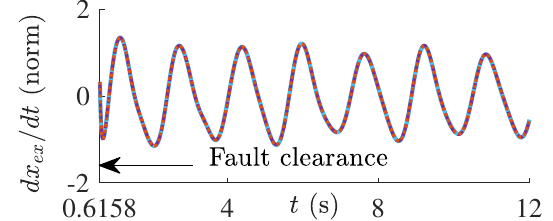}
        \caption{Open-loop training performance}
         \label{fig:dnn:open}
    \end{subfigure}
    \begin{subfigure}{0.47\columnwidth}
        \includegraphics[width=\columnwidth]{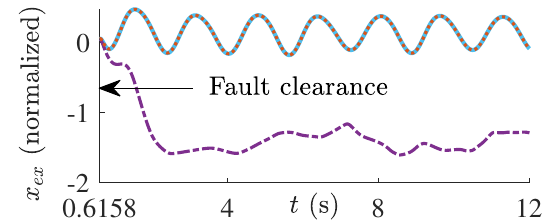}
        \caption{Closed-loop testing performance}
        \label{fig:dnn:close}
    \end{subfigure}
    \vspace{-3pt}
    \caption{Comparison of \neudyn with conventional discrete-time DNN. } \label{fig:dnn}
    \vspace{-10pt}
\end{figure}

Fig.~\ref{fig:dnn} compares \neudyn with conventional deep neural network (DNN) methods to demonstrate the necessity of developing continuous-time learning-based \neudyn. 
Fig.~\ref{fig:dnn:close} shows that DNN fails to provide qualified dynamic responses in the closed-loop simulation despite a perfect training accuracy in Fig.~\ref{fig:dnn:open}.
The fundamental reason is that conventional DNN methods can not directly handle the continuous-time differential equations in \eqref{equ:pi:model}. Therefore, they usually rely on discrete-time learning to obtain a \textit{discretized} dynamic equivalence, whose performance is very sensitive to the training errors. 
%

 \vspace{-2pt}
\subsubsection{Necessity of the physics-aware consideration}
\begin{figure}[!ht]
\vspace{-10pt}
    \centering
     \begin{subfigure}{0.47\columnwidth}
        \includegraphics[width=\columnwidth]{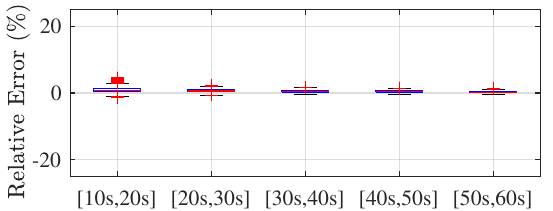}
        \caption{Error of \pineudyn}
         \label{fig:PI error box:PI}
    \end{subfigure}
    \begin{subfigure}{0.47\columnwidth}
        \includegraphics[width=\columnwidth]{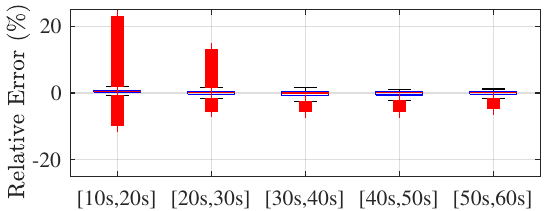}
        \caption{Error of purely data-driven method}
        \label{fig:PI error box:odenet}
    \end{subfigure}
    \vspace{-3pt}
     \caption{\pineudyn's error of tie-line currents under 10 fault cases.}
    \label{fig:PI error box}
    \vspace{-10pt}
\end{figure}

Fig.~\ref{fig:PI error box} compares our physics-aware \neudyn with purely data-driven methods to demonstrate the necessity of incorporating physics knowledge into \neudyn.
The relatively large error in Fig.~\ref{fig:PI error box:odenet} indicates the purely data-driven dynamic equivalence does not guarantee its closed-loop accuracy even though the model is well trained. 
This observation aligns with the discussion in Subsection~\ref{sec:odenet} because purely data-driven methods can not explicitly control the closed-loop accuracy in their training.  
In contrast, physics-aware \neudyn actively incorporates the bi-directional interactions between \insys and \exsys, and hence supports controllable closed-loop accuracy for \neudyn. 

\vspace{-10pt}
\section{Conclusion}
This letter devises physics-aware Neural Dynamic Equivalence (\neudyn), a novel technique to discover a \textit{continuous-time} dynamic equivalence of external systems leveraging \textit{physics knowledge} of internal systems.  
The most salient feature is its capability of retaining the continuous-time dynamic natures  of power grids and its active control of the closed-loop accuracy during training. 
Case studies in the NPCC system show the efficacy of \neudyn under various fault locations, fault clearing times, parameter change, and its superiority over existing discrete-time learning or purely data-driven methods.  

\vspace{-10pt}
\bibliographystyle{ieeetr}
\bibliography{ref}
\end{document}